# Temperature and Thickness Dependent Spin Current Transmission through Epitaxial Antiferromagnetic α-$Fe_2O_3$

**Koushik Prabhakara, Rajeev Nepal, Subhashree Chatterjee, Rabindra Basnet, Ezana Negusse and Ramesh C Budhani[$].**

**Department of Physics, Morgan State University**

**1700 E cold Spring Lane, Baltimore – MD 21251, USA.**

**Abstract:**

Antiferromagnetic insulator α-$Fe_2O_3$ is a potential medium for coherent motion of spin waves over relatively large distances, allowing non-dissipative transfer of information between circuit elements. However, the structural defects and temperature dependent spin reorientations in this oxide impact the propagation. Understanding how the magnetic state of an antiferromagnetic insulator affects spin-current transmission is therefore important for elucidating the mechanisms governing spin transport. Here, we report spin-current transmission through epitaxial α-$Fe_2O_3$ thin films sandwiched between Permalloy and Pt. Using spin pumping and inverse spin Hall effect, we observe a finite symmetric dc voltage across the Pt layer over the temperature range 1.6 to 300 K for α-$Fe_2O_3$ thicknesses of 0, 30, 40, and 100 nm. The symmetric voltage coefficient $K_s$ exhibits two approximately linear temperature regimes, separated by a characteristic temperature $T^*$ whose value depends on the thickness of the α-$Fe_2O_3$ layer. Further, the dc magnetization measurements on some of the samples suggest that the $T^*$ is correlated with the spin reorientation transition (Morin transition) in α-$Fe_2O_3$. Interestingly, a significantly larger magnitude of $dK_s/dT$ is noted below the characteristic temperature $T^*$. The observed thickness and temperature dependent spin-current transmission indicates a sensitivity to the magnetic state of the insulating hematite layer.

$ Ramesh.budhani@morgan.edu

The ability to generate, transport, and manipulate spin angular momentum without an accompanying charge current is a central objective of minimally dissipative spintronics. Magnetic insulators are particularly attractive for this purpose because their collective magnetic excitations can propagate over large distances without loss of coherence[1]. Ferrimagnetic insulators like yttrium iron garnet have provided a platform for studying spin transport[2–4], while antiferromagnetic (AFM) insulators offer additional possibilities associated with their compensated magnetic structure, high-frequency dynamics, and negligible stray fields. Experiments on oxide-based AFMs like NiO, $Cr_2O_3$ and $Fe_2O_3$ have demonstrated spin-angular-momentum transmission through nanometre-thick layers and subsequent detection in an adjacent heavy metal, establishing AFM insulators as potential media for spin transport[5–9].

Spin pumping combined with inverse spin Hall effect (ISHE) in a strongly spin-orbit-coupled (SOC) metal like Pt provides a convenient means of investigating spin transmission across insulating layers. During ferromagnetic resonance (FMR), a driven ferromagnet can pump spin current into an adjacent material[10]. When a SOC heavy metal is separated from the ferromagnet by an insulating magnetic layer, the transmitted spin current can be detected in the heavy metal through its ISHE response. Studies of YIG/NiO/Pt structures have shown that spin-pumping signal depends on AFM-layer thickness, while experiments on epitaxial NiO have demonstrated the importance of crystallographic orientation and interfacial properties[5,11,12]. These results indicate that spin-current transmission is sensitive not only to magnetic properties of the insulating layer but also to its structural and interfacial characteristics.

Hematite, $\alpha$-$Fe_2O_3$ is a particularly interesting AFM-insulating system for spin pumping and spin propagation studies. It has a high Néel temperature (~955 K) and undergoes a Morin transition between different magnetic configurations. Across the transition, the Fe moments reorient from *ab* plane to *c* axis, providing an opportunity to examine how spin transmission evolves with the magnetic state of the antiferromagnet[13–16]. Spin transport through crystalline hematite has been observed over extended distances and across Morin transition[17]. Recent studies have also demonstrated magnon transport and magnon torque in epitaxial $\alpha$-$Fe_2O_3$ films[18,19]. Importantly, the Morin transition in hematite thin films is sensitive to thickness, strain, structural quality, substrate-induced effects, and magnetic field, and can therefore occur at temperatures substantially different from the bulk value[13,15,20]. Such sensitivity raises a fundamental question of how does spin-current transmission change when the magnetic state of an epitaxial hematite layer is altered? Therefore, temperature dependent measurements across this transition are essential for establishing the relationship between the magnetic state of hematite and spin-current transmission.

Although spin pumping and ISHE have been demonstrated in $\alpha$-$Fe_2O_3$/Pt heterostructures, these experiments primarily probe spin pumping generated by the magnetic dynamics of hematite itself[17,21]. Additionally, most of these studies have been conducted at ambient temperatures[19,22,23]. However, to the best of our knowledge, the effects of the temperature and thickness-dependent magnetic state of epitaxial $\alpha$-$Fe_2O_3$ on its spin current transmission characteristics have not been reported. Here, we investigate this phenomenon in epitaxial $\alpha$-$Fe_2O_3$ layers sandwiched between the films of permalloy (Py) and Pt. We study heterostructures containing 30, 40, and 100 nm thick α-$Fe_2O_3$ layers, together with a Py/Pt reference sample, over a wide temperature range spanning from 1.6 to 300 K in an ISHE/FMR spectrometer. These experiments reveal a dc voltage across the Pt layer associated with spin-current transmission through the hematite over entire temperature range investigated. The temperature dependence of symmetric ISHE voltage coefficient, $K_s$ exhibits distinct low and high-temperature regimes separated by a characteristic $T^*$ which depends on the hematite thickness. This behaviour of $K_s$ indicates a connection between spin-current transmission and reorientation of Fe-moments from *ab*-plane to along the *c*-axis at the Morin transition in hexagonal $Fe_2O_3$.

Thin-film heterostructures of the type (0001) $Al_2O_3$/Pt (10 nm)/α-$Fe_2O_3$($x$ nm)/Py (30 nm), with $x$ = 30, 40 and 100 nm were deposited using a load-lock, multi-target high-vacuum magnetron sputtering system with a base pressure of 8 x $10^{-9}$ Torr. Platinum is known to grow epitaxially with (111) orientation on *C*-plane sapphire due to low lattice mismatch of 0.9%[24]. Pt seed layer was deposited at 500 ºC, followed by cooling to 350 ºC for deposition of subsequent Pt and α-$Fe_2O_3$ layers. The initial high-temperature Pt deposition promotes epitaxial Pt (111) nucleation, while subsequent lower-temperature deposition suppresses 3-dimensional island growth and surface roughness associated with prolonged high-temperature Pt growth[24]. The α-$Fe_2O_3$ layers were deposited by reactive sputtering of Fe in an Ar-$O_2$ atmosphere with Ar:$O_2$ flow of ~ 12.5:4 sccm. The Py layer was subsequently deposited in pure argon at room temperature after flushing out oxygen, followed by an Al capping layer to protect the heterostructure from oxidation. The crystallographic structure of the stack was evaluated by X-ray diffraction measurements performed using a Rigaku SmartLab diffractometer. Surface morphology and magnetic textures of the films were evaluated using Park Systems NX10 scanning probe microscope. For in-plane magnetization measurements, we have used the vibrating sample magnetometer (VSM) option of physical property measurement system equipped with a 9 T superconducting magnet. FMR measurements were performed using a custom-built broadband Cryo-FMR spectrometer operated in frequency-modulation mode from 1.6 to 300 K[25]. The Py/α-$Fe_2O_3$/Pt/(0001) $Al_2O_3$ samples were placed face down on the signal line of grounded coplanar waveguide (GCPW) inside variable-temperature insert of a superconducting magnet. A frequency-modulated microwave signal was applied to GCPW, and transmitted signal was rectified and detected using a lock-in amplifier. Mutually perpendicular dc and RF magnetic fields were oriented in the sample plane. Simultaneously, FMR-induced dc voltage, including ISHE and spin-rectification contributions, was measured using a nanovoltmeter.

We first characterize structural and magnetic properties of the heterostructures used for spin-current transmission measurements. The AFM images in Fig. 1(a) and (b) show a strong dependence of Pt morphology on the oxygen content in the plasma during sputtering. Without oxygen (Fig. 1(a)), the 9-nm Pt film exhibits a discontinuous, maze-like morphology with irregular islands, indicative of Volmer-Weber growth[26], with a surface roughness of ~2.1 nm. This suggests limited wetting of Pt on $Al_2O_3$ and high surface diffusion, leading to isolated clusters. In contrast, Pt deposited in oxygen (Fig. 1(b)), shows elongated, more continuous features with reduced surface roughness of ~1.7 nm, indicating improved film continuity due to modified surface interactions. The morphology of a 50 nm thick $Fe_2O_3$ film grown on Pt (111) is shown in Fig. 1(c), with a grain size of ~80 nm and surface roughness of ~1.2 nm. Fig. 1(d) shows the $2\theta$-$\omega$ X-ray diffraction pattern of the heterostructure grown on *c*-plane sapphire. Two prominent features are observed: the $Al_2O_3$ (0006) reflection appearing at higher $2\theta$ and a feature near ~ 40º. The latter feature is associated with the Pt (111) and α-$Fe_2O_3$ (006) reflections. Since Pt (111) and α-$Fe_2O_3$ (006) reflections occur at closely spaced 2θ positions, the observed reflection cannot be unambiguously assigned to α-$Fe_2O_3$ (0006) alone. We therefore attribute this feature to the combined diffraction contribution from the Pt (111) and α-$Fe_2O_3$ (006) planes which is consistent with previous studies[19]. Fig. 2(a) shows the temperature dependence of normalized magnetization of α-$Fe_2O_3$/Pt heterostructure measured under zero-field-cooled (black) and field-cooled (red) conditions. As the temperature decreases, magnetization gradually reduces and reaches a minimum near $T$ ~ 80 K. Since the hematite at low-temperatures is a fully compensated antiferromagnet with the Fe moments aligned along the c-axis[17], this behaviour provides a possible indication of a Morin transition near $T_M$ ~ 80 K. This value is substantially lower than the bulk Morin transition temperature (~260 K), but such a reduction is plausible in thin films, where $T_M$ depends on film thickness, lattice strain, structural defects, and interfacial conditions[13,15,20]. It is important to examine whether the change in magnetic configuration influences the transmission of spin angular momentum. Hence, we employ FMR-driven spin pumping from Py layer and detect resulting voltage in Pt layer. The measurement geometry is illustrated schematically in the inset of Fig. 1(d).

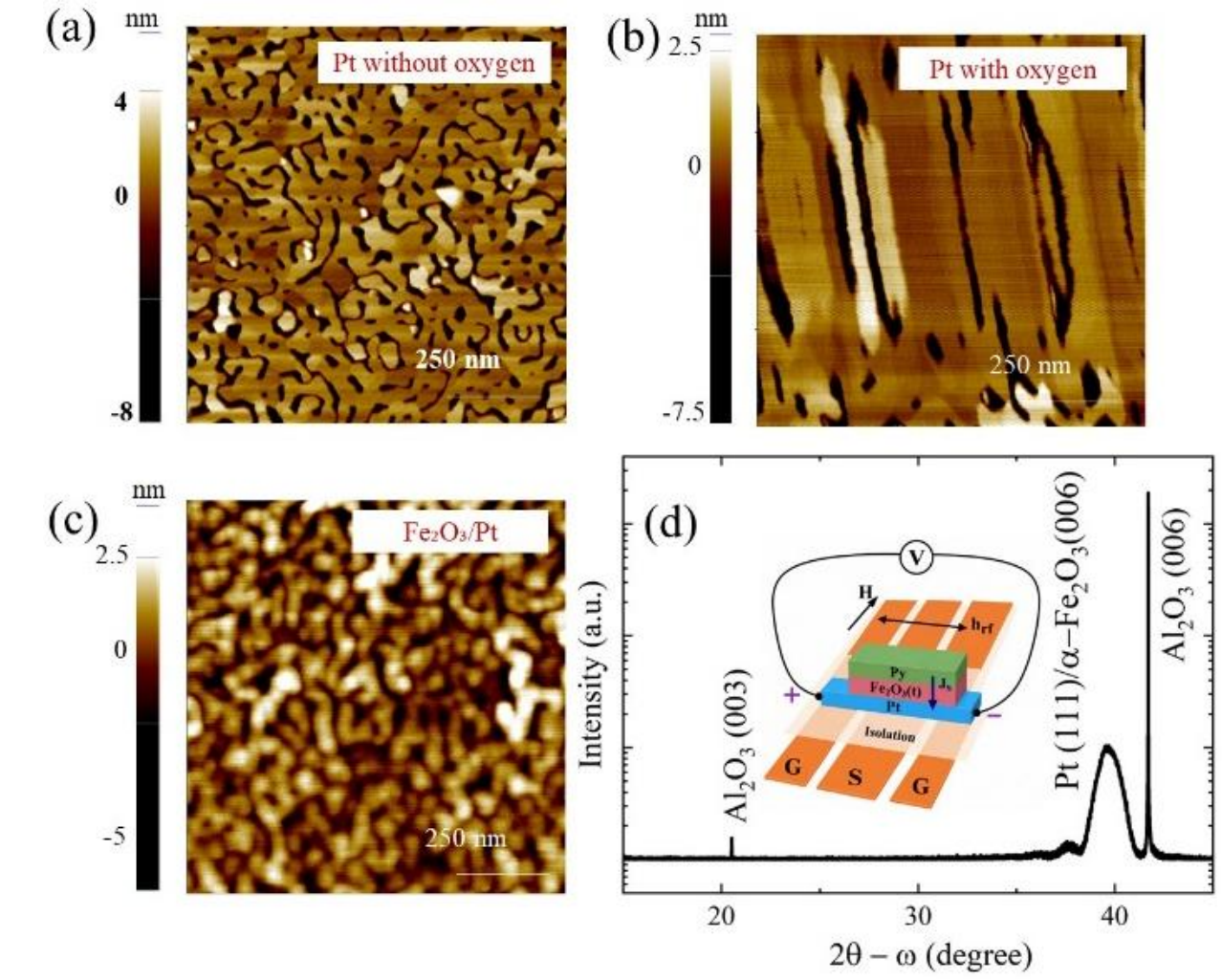

***Figure 1:*** *AFM surface topography of (a) 9 nm Pt deposited without oxygen, (b) 9 nm Pt deposited in oxygen, (c) 50 nm $Fe_2O_3$ grown on a 9 nm Pt-coated $Al_2O_3$ substrate and (d) XRD θ–2θ pattern of $Fe_2O_3$/Pt on (0001) $Al_2O_3$. Inset: Schematic of used for spin-current transmission measurements.*

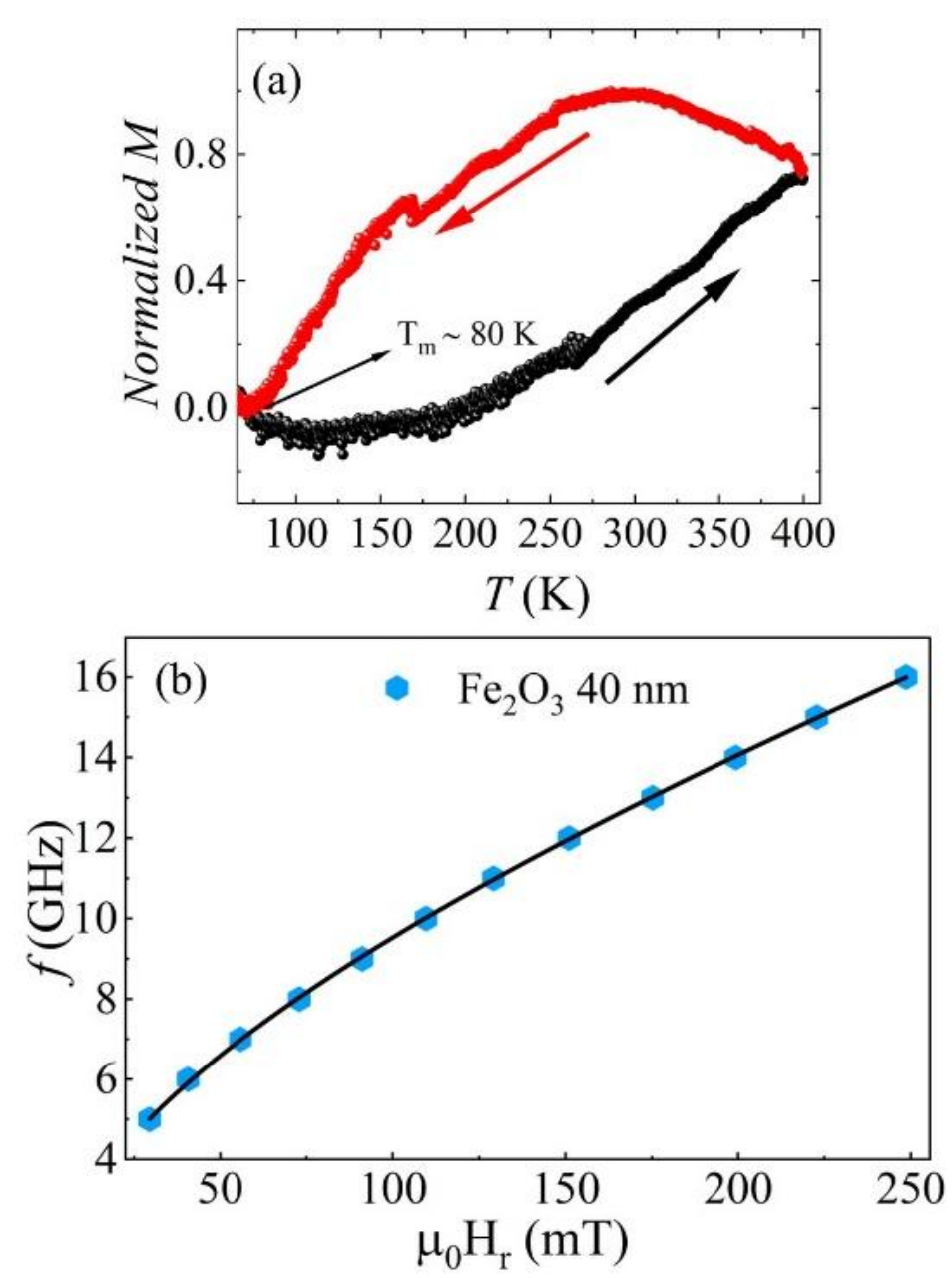

***Figure 2:*** ***(a)*** *Normalized magnetization of Pt/α-$Fe_2O_3$(100 nm) under zero-field-cooled (black) and field-cooled (red) conditions.* ***(b)*** *Resonance frequency versus resonance field for Pt/α-$Fe_2O_3$ (40 nm)/Py heterostructure. The solid line is a fit to Kittel expression.*

During FMR excitation, spin current $J_s$ is pumped from the precessing Py magnetization toward Pt through the intervening α-$Fe_2O_3$. The transmitted spin current is converted into a transverse dc charge current via ISHE. FMR-induced voltages can also contain spin-rectification contributions from the ferromagnetic layer[27–29]. To minimize the latter effect, the voltage contacts were made only on the Pt layer (see inset of Fig. 1d). The measured voltage is subsequently decomposed into its symmetric and antisymmetric components, allowing the symmetric contribution associated with spin-current transmission to be extracted for further analysis[25]. The characteristic resonance fields $H_r$ were obtained by fitting the field-dependent $V_{mix}$ signals and used to determine the frequency dependence of the resonance. Fig. 2 (b) presents $f_r$ as a function of corresponding $H_r$ for Py/α-$Fe_2O_3$(40 nm)/Pt heterostructure. The observed trend is well described by Kittel relation;

$$f_r = \frac{\gamma}{2\pi}\mu_0\sqrt{H_r(H_r + M_{eff})} \ldots\ldots (1)$$

where $\gamma = g\mu_B/h$. Here $g$, $\mu_B$ and $h$ are Lande's g factor, Bohr magneton and Planck's constant, respectively and $M_{eff}$ is the effective magnetization[30,31]. The fit yields g = 2.03 ± 0.04 and $M_{eff}$ = 1.02 ± 0.05 T, consistent with permalloy thin films[32,33]. This agreement establishes the basis for the subsequent measurements of the FMR-induced voltage and spin-current transmission. We next separate symmetric and antisymmetric components of $V_{mix}$ to identify the contributions associated with spin-current transmission through the α-$Fe_2O_3$ layer. Fig. 3 shows representative magnetic-field-dependent $V_{mix}$ signals measured from 1.6 to 300 K for the Py/Pt reference and for Py/α-$Fe_2O_3$ (t = 30, 40 and 100 nm)/Pt heterostructures.

The overall line shape and amplitude of $V_{mix}$ response remains observable over the entire temperature range, providing a reference for the voltage response in the absence of α-$Fe_2O_3$ layer. Upon insertion of the hematite layer, a clear voltage response remains detectable for all three thicknesses, from 1.6 K to room temperature, as shown in figs. 3 (b-d).

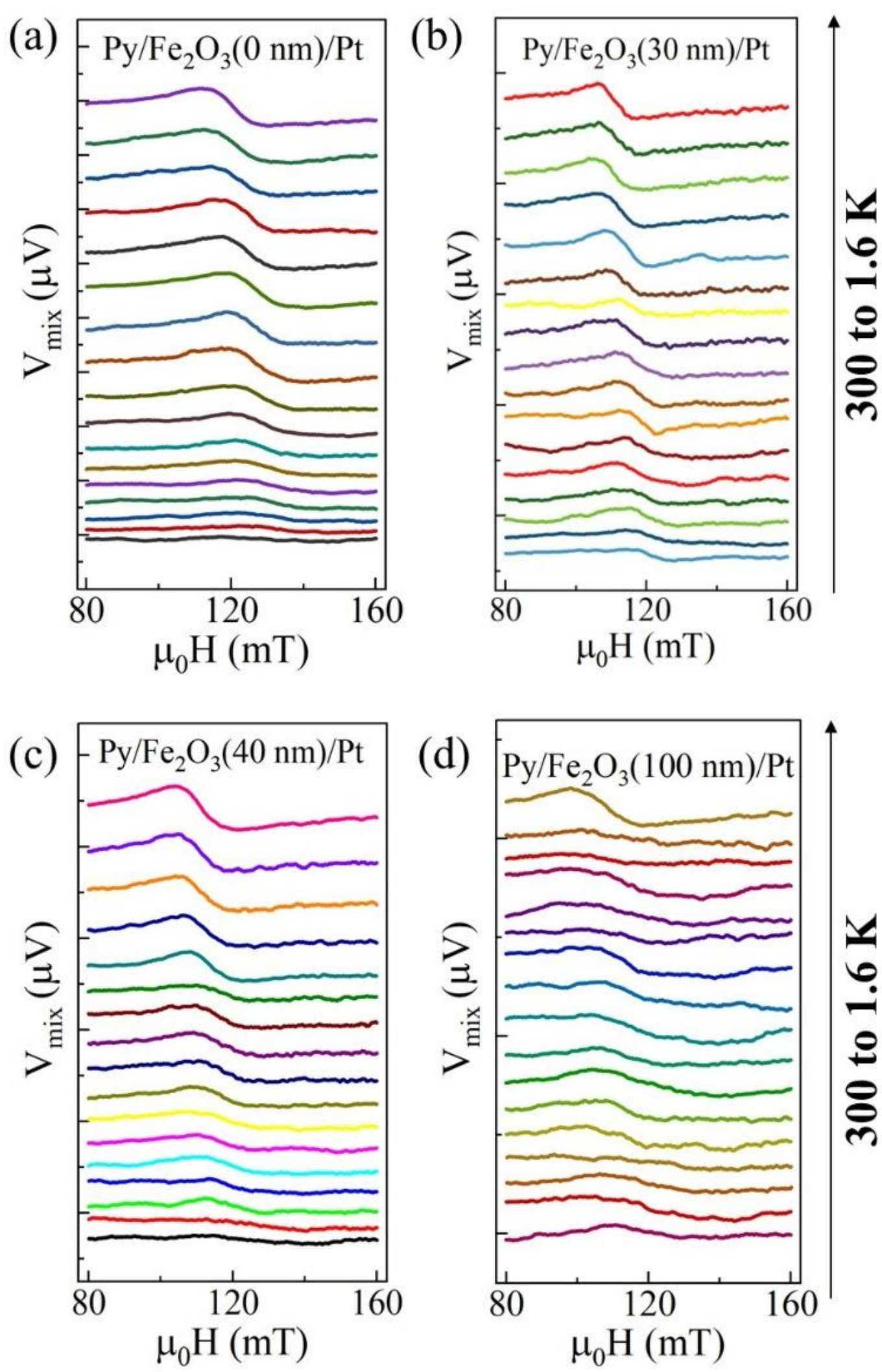

***Figure 3:*** *Field dependence of $V_{mix}$ for Py/α-$Fe_2O_3$/Pt heterostructures with α-$Fe_2O_3$ thicknesses of **(a)** 0, **(b)** 30, **(c)** 40, and **(d)** 100 nm over the temperature range 1.6 – 300 K, in 20 K steps. Curves are vertically offset for clarity.*

The persistence of voltage after insertion of α-$Fe_2O_3$ is important because the insulating layer electrically separates the Py injector from the Pt detector. Thus, the measured response cannot be attributed to direct charge conduction through α-$Fe_2O_3$ layer. Similar FMR-based experiments on insulating AFM spacers have used voltage detection in Pt to probe spin-current transmission[5,11]. However, $V_{mix}$ signal may be contaminated by contributions unrelated to spin pumping and the resultant ISHE voltage[28,34]. A collective effect of such signals is to make the $V_{mix}$ asymmetric in field. The antisymmetric components can arise from anisotropic magnetoresistance and anomalous Hall effect in Py, together with thermal-gradient contributions, whereas the ISHE contribution exhibits a symmetric Lorentzian line shape[35–37]. We therefore separately analyse symmetric and antisymmetric parts of $V_{mix}$.

Figs. 4 (a) and (b) show representative field-dependent voltage measurements for Py/Pt reference bilayer and Py/α-$Fe_2O_3$(100 nm)/Pt heterostructure at 1.6 K. Both samples exhibit pronounced voltage response near the FMR field. The measured voltage is decomposed into symmetric and antisymmetric components using;

$$V_{mix}(H) = K_s \frac{\Delta H^2}{(H-H_r)^2+\Delta H^2} + K_{as} \frac{-2\Delta H(H-H_r)}{(H-H_r)^2+\Delta H^2} \dots (2)$$

Where $K_s$ and $K_{as}$ are amplitudes of the symmetric and antisymmetric components, respectively, and $\Delta H$ is the half width at half maximum of $V_{mix}$ $(H)$[38–40]. The black curves in Figs. 4(a) and 4(b) represent total fits, with corresponding symmetric components shown in insets. For the Py/Pt reference, symmetric component reaches ~ 0.39 μV, while ~ 0.21 μV is obtained for the Py/α-$Fe_2O_3$(100 nm)/Pt heterostructure.

The persistence of the symmetric voltage after insertion of 100 nm α-$Fe_2O_3$ layer is consistent with spin-current transmission from Py to Pt. We associate $K_s$ with the spin current related voltage detected in Pt, while considering that a FMR induced voltage can also contain contributions from spin rectification or thermoelectric effects[29,38,41,42].

We examine temperature and thickness dependence of $K_s$ below. Figs. 5 (a–d) show $K_s$ as a function of temperature for the Py/Pt reference and for the Py/α-$Fe_2O_3$/Pt heterostructures with α-$Fe_2O_3$ thicknesses of 30, 40, and 100 nm, respectively. The Py/Pt bilayer exhibits a nearly linear decrease of $K_s$ with increasing temperature, with $dK_s/dT$ = (-7.24 ± 0.18) x $10^{-10}$ V/K.

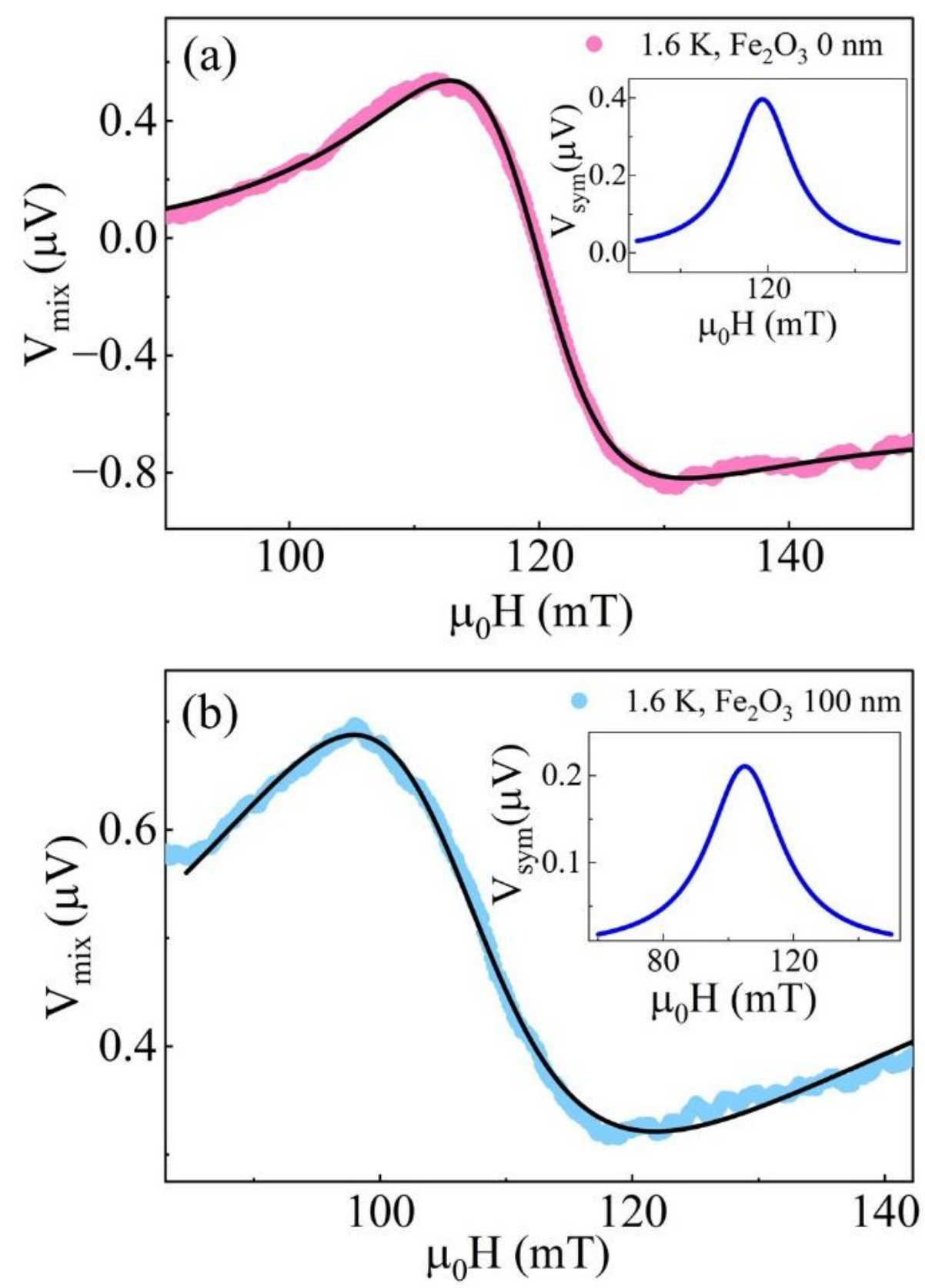

***Figure 4:*** *Field dependence of $V_{mix}$ for **(a)** Py/Pt and (b) α-$Fe_2O_3$ 100 nm at 1.6 K, respectively. Solid lines are fits using Eq. 2. Insets show the corresponding symmetric Lorentzian components.*

In contrast, $\alpha$-$Fe_2O_3$ containing samples show a pronounced change in the slope of $K_s$ at low temperatures. For 30 and 40 nm $\alpha$-$Fe_2O_3$ films, $K_s$ decreases more rapidly below $T^* \sim 40$ K and exhibits a substantially weaker temperature dependence above this characteristic temperature. A similar change occurs for 100 nm film, although $T^*$ shifts to ~ 80 K. The dashed lines in Fig. 5 indicate the change in slope near $T^*$. The linear fits to the two regimes quantify the corresponding $dK_s/dT$ and the values are summarized in Table 1. For all three $\alpha$-$Fe_2O_3$ thicknesses, $K_s$ decreases more rapidly below $T^*$ as indicated by the larger magnitude of $dK_s/dT$ whereas its temperature dependence is weaker above $T^*$.

These results demonstrate that the temperature dependence of $K_s$ is strongly modified by the presence and thickness of the $\alpha$-$Fe_2O_3$ layer in these heterostructures. The characteristic temperatures extracted from the behaviour of $K_s$ can be compared with the magnetic response.

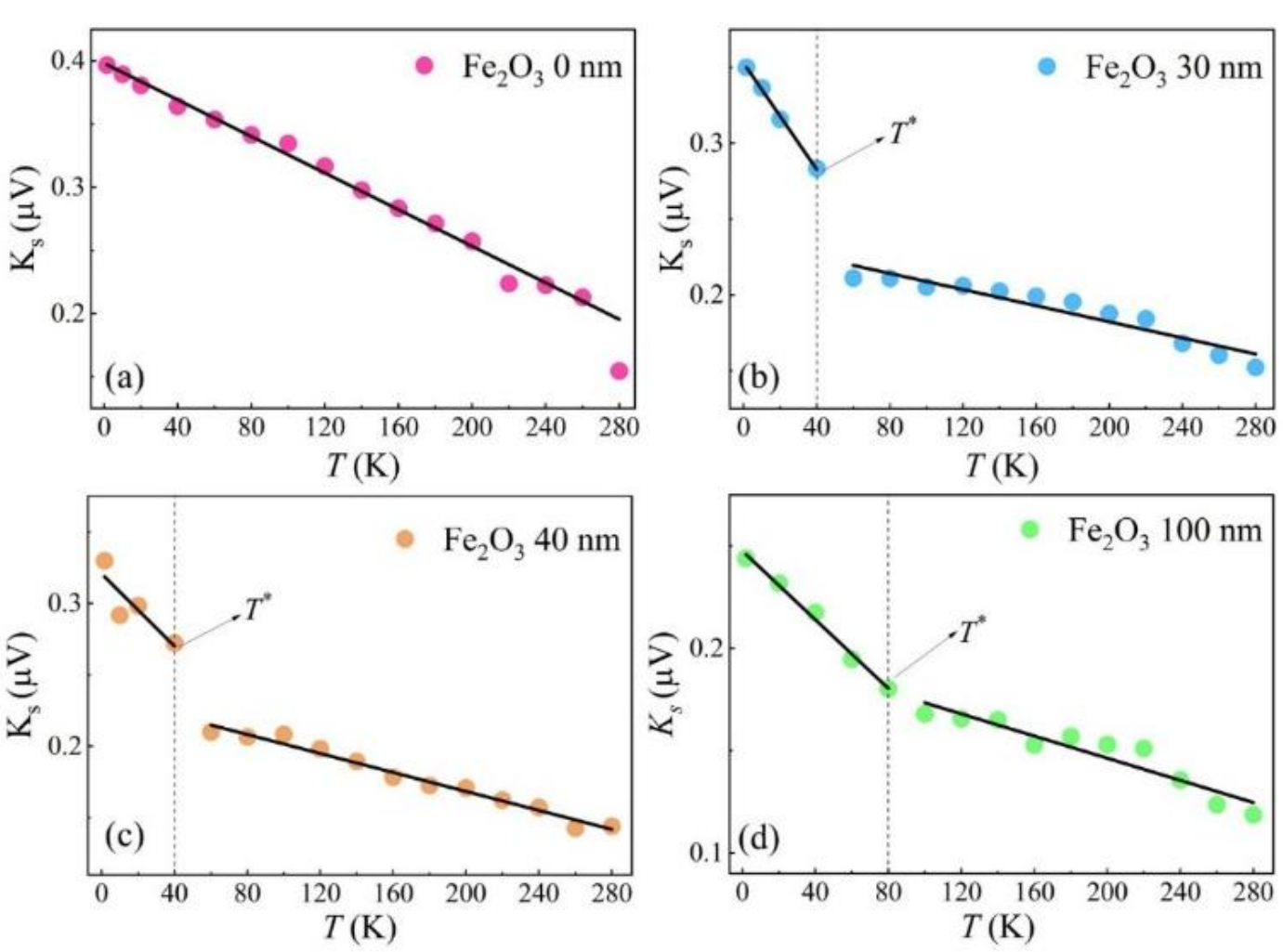

***Figure 5:*** *Temperature dependence of $K_s$ for Pt/$\alpha$-$Fe_2O_3$/Py heterostructures with $\alpha$-$Fe_2O_3$ thicknesses of* ***(a)*** *0,* ***(b)*** *30,* ***(c)*** *40, and* ***(d)*** *100 nm. Solid lines are linear fits over the respective temperature ranges; dashed lines indicate $T^*$.*

$T^* \sim 80$ K observed for the 100-nm $\alpha$-$Fe_2O_3$ film is close to the temperature at which magnetization of the Pt/$\alpha$-$Fe_2O_3$ heterostructure approaches zero. This correspondence suggests that the change in $K_s$ is associated with the magnetic reorientation of $\alpha$-$Fe_2O_3$ in the vicinity of the Morin transition regime. The large drop in $T^*$ for the 30 and 40 nm films appears to be an indication of a significant change in the $T_M$ of these thinner films. As stated earlier, the Morin transition $T_M$ in hematite films depend on thickness, strain, and magnetic field, and it may shift depending on growth conditions and strain[13]. Taken together, these observations suggest that the temperature dependence of spin-current transmission is sensitive to the magnetic state of $\alpha$-$Fe_2O_3$ layer. To further characterize spin-current transfer, we analyse the ISHE resonance linewidth $\Delta H$ within the Landau-Lifshitz-Gilbert model

$$\Delta H = \Delta H_0 + \frac{2\pi}{\gamma}\alpha f \ldots\ldots (3)$$

***Table 1:*** *Linear-fit parameters for $K_s$ (T) of Py/Pt reference and Pt/$\alpha$-$Fe_2O_3$/Py heterostructures.*

| **$\alpha$-$Fe_2O_3$ thickness** | **$T^*$ (K)** | ***$dK_s/dT$ below $T^*$* (V/K)** | ***$dK_s/dT$ above $T^*$* (V/K)** |
|---|---|---|---|
| 0 nm (Py/Pt) | - | - | $-7.24 \pm 0.17 \times 10^{-10}$ |
| 30 nm | ~ 40 | $-1.75 \pm 0.05 \times 10^{-9}$ | $-2.66 \pm 0.27 \times 10^{-10}$ |
| 40 nm | ~ 40 | $-1.26 \pm 0.49 \times 10^{-9}$ | $-3.32 \pm 0.16 \times 10^{-10}$ |
| 100 nm | ~ 80 | $-1.13 \pm 0.15 \times 10^{-9}$ | $-8.40 \pm 0.45 \times 10^{-10}$ |

Where $\Delta H_0$ is the zero-field linewidth resulting from frozen-in defects in materials and $\alpha$ is Gilbert damping constant[43]. The fitted values of $\alpha$ are found to be 6.21 x $10^{-3}$, 6.59 x $10^{-3}$, 6.58 x $10^{-3}$ for Py/Pt reference, 30, and 40 nm $Fe_2O_3$ films, respectively. The enhancement in damping relative to the Py/Pt reference is indicative of spin pumping that allows transfer of spin angular momentum from Py to $Fe_2O_3$ layer. Further, the spin pumping efficiency of interface can be evaluated by calculating real part of the spin mixing conductance $g^{\uparrow\downarrow}$ using the following relation;

$$g^{\uparrow\downarrow} = \frac{4\pi M_s t_{Py}}{g\mu_B}\left(\alpha_{Py/Fe_2O_3(t)/Pt} - \alpha_{Py/Pt}\right) \ldots\ldots (4)$$

Where $M_s$ and $t_{Py}$ are the saturation magnetization and thickness of the Py layer, respectively[34,44]. Using $\mu_0 M_s$ = 1 T[44,45] and $t_{py}$ = 30 nm, we obtain $g_{eff}^{\uparrow\downarrow} \sim 6.15 \times 10^{18}$ and $5.98 \times 10^{18}$ $m^{-2}$ for 30 and 40 nm α-$Fe_2O_3$ layers, respectively. The reduction in $g_{eff}^{\uparrow\downarrow}$ with increasing α-$Fe_2O_3$ thickness is consistent with reduced spin-current transfer through the spacer.

We note that the prominent feature of our measurements is not simply the presence of transmitted spin current but change in its temperature dependence. As shown in fig. 5, $K_s$ for all three α-$Fe_2O_3$ thicknesses exhibits two linear regimes with different slopes. The magnitude of $dK_s/dT$ is larger at lower temperatures, while characteristic temperature $T^*$ depends on α-$Fe_2O_3$ thickness as seen from its value of ~ 40 K for the 30 and 40 nm films and ~ 80 K for 100 nm film. The latter is comparable to the temperature range in which the magnetization of Pt/α-$Fe_2O_3$ sample becomes nearly zero, suggesting that change in $K_s$ is associated with magnetic reorientation regime of α-$Fe_2O_3$. This interpretation is consistent with previous observations that spin transport depends on the magnetic state and thickness of antiferromagnetic insulators[46]. Such sensitivity is not unexpected in epitaxial antiferromagnetic films, where strain, crystallinity, interfaces, and magnetic configuration can influence spin dependent properties[13,20]. Hence, these results motivate experiments that directly resolve the magnetic excitations contributing to spin transport. For example, broadband GHz spin-wave spectroscopy or Brillouin light scattering could determine spin-wave dispersion, damping, and propagation characteristics[23], while time-resolved THz spectroscopy could probe ultrafast spin-current dynamics and distinguish contributions from coherent antiferromagnetic excitations and spin-caloritronic processes[47].

In summary, we investigated spin-current transmission through epitaxial α-$Fe_2O_3$ thin films incorporated between Py and Pt. The observation of a symmetric dc voltage related to ISHE in Pt demonstrates spin-current transmission through the insulating α-$Fe_2O_3$ spacer over the temperature range 1.6 to 300 K. The temperature dependence of the spin pumping coefficient $K_s$ varies systematically with the thickness of α-$Fe_2O_3$ layer, showing a larger $dK_s/dT$ below a thickness dependent characteristic temperature. The correspondence between $T^*$ and the magnetic response of the Pt/α-$Fe_2O_3$ sample suggests that this change is associated with magnetic reorientation of α-$Fe_2O_3$. These results motivate further studies across the Morin-transition regime in epitaxial hematite.

**Acknowledgements**

This work has been supported in part by the National Science Foundation through grant No. 2309672. Part of this work was also supported by the Center for Energy Efficient Magnonics, an Energy Frontier Research Center funded by the U.S. Department of Energy, Office of Science, Basic Energy Sciences, under Contract No. DE-AC02-76SF00515.

## References

[1] A. Brataas, B. van Wees, O. Klein, G. de Loubens, and M. Viret, "Spin insulatronics," Phys. Rep. **885**, 1 (2020).

[2] D. Wesenberg, T. Liu, D. Balzar, M. Wu, and B.L. Zink, "Long-distance spin transport in a disordered magnetic insulator," Nat. Phys. **13**, 987 (2017).

[3] K.S. Rathore, A. Swain, J. Lim, A. Mishra, P. Gupta, L. Yong Heng, J. Luwen, R. Mahendiran, A. Hoffmann, and S. Bedanta, "Magnon-magnon coupling in an all-oxide insulator garnet $Y_3Fe_5O_{12}/Tm_3Fe_5O_{12}$ heterostructure," Phys. Rev. Appl. **25**, 024036 (2026).

[4] J. Li, Y. Xu, M. Aldosary, C. Tang, Z. Lin, S. Zhang, R. Lake, and J. Shi, "Observation of magnon-mediated current drag in Pt/yttrium iron garnet/Pt(Ta) trilayers," Nat. Commun. **7**, 10858 (2016).

[5] H. Wang, C. Du, P.C. Hammel, and F. Yang, "Antiferromagnonic Spin Transport from $Y_3Fe_5O_{12}$ into NiO," Phys. Rev. Lett. **113**, 097202 (2014).

[6] Y. Pu, G. Shi, C. Zhang, X. Chen, H. Park, and H. Yang, "Crystalline-Dependent Magnon Torques in All-Sputtered Hf/$Cr_2O_3$/Ferromagnet Heterostructures," Advanced Materials **37**, 2501989 (2025).

[7] Z. Qiu, D. Hou, J. Barker, K. Yamamoto, O. Gomonay, and E. Saitoh, "Spin colossal magnetoresistance in an antiferromagnetic insulator," Nat. Mater. **17**, 577 (2018).

[8] A. Zou, Z. Zhang, Y. Xu, M. Du, X. Wang, S. Chang, and Z. Qiu, "Study of interfacial spin transport with an antiferromagnetic $Cr_2O_3$ interlayer," Phys. Rev. B **105**, 014427 (2022).

[9] G.R. Hoogeboom, G.-J.N.S. Nicolaas, A. Alexander, O. Kuschel, J. Wollschläger, I. Ennen, B.J. van Wees, and T. Kuschel, "Role of NiO in the nonlocal spin transport through thin NiO films on $Y_3Fe_5O_{12}$," Phys. Rev. B **103**, 144406 (2021).

[10] Y. Wang, R. Ramaswamy, and H. Yang, "FMR-related phenomena in spintronic devices," J. Phys. D Appl. Phys. **51**, 273002 (2018).

[11] L. Jin, K. Jia, D. Zhang, B. Liu, H. Meng, X. Tang, Z. Zhong, and H. Zhang, "Effect of Interfacial Roughness Spin Scattering on the Spin Current Transport in YIG/NiO/Pt Heterostructures," ACS Appl. Mater. Interfaces **11**, 35458 (2019).

[12] S. Chen, A. Pomar, L. Balcells, Z. Konstantinovic, B. Bozzo, C. Frontera, C. Magen, N. Mestres, and B. Martinez, "Crystallographic Engineering of Spin Transport in Antiferromagnetic NiO Thin Films," ACS Nano **19**, 32170(2025).

[13] S. Park, H. Jang, J.-Y. Kim, B.-G. Park, T.-Y. Koo, and J.-H. Park, "Strain control of Morin temperature in epitaxial α-$Fe_2O_3$ (0001) film," Europhysics Letters **103**, 27007 (2013).

[14] N. Shimomura, S.P. Pati, Y. Sato, T. Nozaki, T. Shibata, K. Mibu, and M. Sahashi, "Morin transition temperature in (0001)-oriented α-$Fe_2O_3$ thin film and effect of Ir doping," J. Appl. Phys. **117**, 17C736 (2015).

[15] L. Huang, C.F. Li, Y.S. Tang, L. Lin, W.J. Zhai, X.M. Cui, G.Z. Zhou, J.H. Zhang, Z.B. Yan, C. Chen, X.P. Jiang, C.L. Lu, and J.-M. Liu, "Magnetotransport around the Morin transition in α-$Fe_2O_3$ single crystals," J. Appl. Phys. **132**, 163903 (2022).

[16] J. Lielmezs, and A.C.D. Chaklader, "Magnetic Transitions in Alpha Hematite," Science (1979). **160**, 1137 (1968).

[17] R. Lebrun, A. Ross, O. Gomonay, V. Baltz, U. Ebels, A.-L. Barra, A. Qaiumzadeh, A. Brataas, J. Sinova, and M. Kläui, "Long-distance spin-transport across the Morin phase transition up to room temperature in ultra-low damping single crystals of the antiferromagnet α-$Fe_2O_3$," Nat. Commun. **11**, 6332 (2020).

[18] J. Han, P. Zhang, Z. Bi, Y. Fan, T.S. Safi, J. Xiang, J. Finley, L. Fu, R. Cheng, and L. Liu, "Birefringence-like spin transport via linearly polarized antiferromagnetic magnons," Nat. Nanotechnol. **15**, 563 (2020).

[19] D. Song, F. Meng, M. Sun, H. Ni, J. Zou, Z. Yao, Z. Guo, Y. Zhang, L. Zhang, H. Bai, W. Zhou, and Y. Wang, "Efficient magnon transport and magnon torque through the epitaxial antiferromagnetic insulator α-$Fe_2O_3$," Chinese Physics B **35**, 047505 (2026).

[20] A. Kozioł-Rachwał, N. Kwiatek, W. Skowroński, K. Grochot, J. Kanak, E. Madej, K. Freindl, J. Korecki, and N. Spiridis, "Insight into the structural and magnetotransport properties of epitaxial α−$Fe_2O_3$/Pt(111) heterostructures: Role of the reversed layer sequence," Phys. Rev. B **106**, 104419 (2022).

[21] G. Fritjofson, J. Tang, A. Regmi, J. Hanson-Flores, J. Michel, F. Yang, R. Cheng, and E. Del Barco, "Coherent Spin Pumping Originated from Sub-Terahertz Néel Vector Dynamics in Easy Plane α-$Fe_2O_3$/Pt," Phys. Rev. Lett. **135**, 216704 (2025).

[22] D. Gabrielyan, D. Volkov, E. Kozlova, A. Safin, D. Kalyabin, and S. Nikitov, "Room-temperature spin pumping from canted antiferromagnet α-$Fe_2O_3$," J. Appl. Phys. **136**, 083902 (2024).

[23] M. Hamdi, F. Posva, and D. Grundler, "Spin wave dispersion of ultra-low damping hematite (α-$Fe_2O_3$) at GHz frequencies," Phys. Rev. Mater. **7**, 054407 (2023).

[24] C.G. Torres-Castanedo, D.B. Buchholz, T. Pham, L. Zheng, M. Cheng, V.P. Dravid, M.C. Hersam, and M.J. Bedzyk, "Ultrasmooth Epitaxial Pt Thin Films Grown by Pulsed Laser Deposition," ACS Appl. Mater. Interfaces **16**, 1921 (2024).

[25] V. Sharma, E. Negusse, R. Kumar, and R.C. Budhani, "Ferromagnetic resonance measurement with frequency modulation down to 2 K," Review of Scientific Instruments **95**, 063902 (2024).

[26] G. Abadias, L. Simonot, J.J. Colin, A. Michel, S. Camelio, and D. Babonneau, "Volmer-Weber growth stages of polycrystalline metal films probed by in situ and real-time optical diagnostics," Appl. Phys. Lett. **107**, 183105 (2015).

[27] P. Wang, S.W. Jiang, Z.Z. Luan, L.F. Zhou, H.F. Ding, Y. Zhou, X.D. Tao, and D. Wu, "Spin rectification induced by spin Hall magnetoresistance at room temperature," Appl. Phys. Lett. **109**, 112406 (2016).

[28] L. Bai, P. Hyde, Y.S. Gui, C.-M. Hu, V. Vlaminck, J.E. Pearson, S.D. Bader, and A. Hoffmann, "Universal Method for Separating Spin Pumping from Spin Rectification Voltage of Ferromagnetic Resonance," Phys. Rev. Lett. **111**, 217602 (2013).

[29] K. He, J. Cheng, M. Yang, L. Sun, W. Sun, S. Bedanta, A. Azevedo, B. Miao, and H. Ding, "Spin rectification effects in ferromagnetic metal microstrips induced by anisotropic magnetoresistance, planar Hall effect, and anomalous Hall effect," Phys. Rev. B **106**, 104407 (2022).

[30] M. Farle, "Ferromagnetic resonance of ultrathin metallic layers," Reports on Progress in Physics **61**, 755 (1998).

[31] R.C. Budhani, S.A. Basun, M.E. McConney, D.R. Evans, M.R. Page, and B.M. Howe, "Temperature dependent resonant microwave absorption in perpendicular magnetic anisotropy epitaxial films of a spinel ferrite," J. Appl. Phys. **125**, 243903 (2019).

[32] J.P. Nibarger, R. Lopusnik, and T.J. Silva, "Damping as a function of pulsed field amplitude and bias field in thin film Permalloy," Appl. Phys. Lett. **82**, 2112 (2003).

[33] J.M. Shaw, H.T. Nembach, T.J. Silva, and C.T. Boone, "Precise determination of the spectroscopic g-factor by use of broadband ferromagnetic resonance spectroscopy," J. Appl. Phys. **114**, 243906 (2013).

[34] O. Mosendz, V. Vlaminck, J.E. Pearson, F.Y. Fradin, G.E.W. Bauer, S.D. Bader, and A. Hoffmann, "Detection and quantification of inverse spin Hall effect from spin pumping

in permalloy/normal metal bilayers," Phys. Rev. B **82**, 214403 (2010).

[35] J. Sklenar, W. Zhang, M.B. Jungfleisch, W. Jiang, H. Saglam, J.E. Pearson, J.B. Ketterson, and A. Hoffmann, "Spin Hall effects in metallic antiferromagnets – perspectives for future spin-orbitronics," AIP Adv. **6**, 055603 (2016).

[36] W. Zhang, M.B. Jungfleisch, W. Jiang, J. Sklenar, F.Y. Fradin, J.E. Pearson, J.B. Ketterson, and A. Hoffmann, "Spin pumping and inverse spin Hall Effects-Insights for future spin-orbitronics (invited)," J. Appl. Phys. **117**, 172610 (2015).

[37] A. Azevedo, L.H. Vilela-Leão, R.L. Rodríguez-Suárez, A.F. Lacerda Santos, and S.M. Rezende, "Spin pumping and anisotropic magnetoresistance voltages in magnetic bilayers: Theory and experiment," Phys. Rev. B **83**, 144402 (2011).

[38] M. Harder, Y. Gui, and C.-M. Hu, "Electrical detection of magnetization dynamics via spin rectification effects," Phys. Rep. **661**, 1 (2016).

[39] P. Bajracharya, V. Sharma, A. Johnson, and R.C. Budhani, "Resonant precession of magnetization and precession—induced DC voltages in FeGaB thin films," J. Phys. D Appl. Phys. **55**, 075303 (2022).

[40] V. Sharma, W. Wu, P. Bajracharya, D.Q. To, A. Johnson, A. Janotti, G.W. Bryant, L. Gundlach, M.B. Jungfleisch, and R.C. Budhani, "Light and microwave driven spin pumping across FeGaB–BiSb interface," Phys. Rev. Mater. **5**, 124410 (2021).

[41] S. Obinata, R. Iimori, K. Ohnishi, and T. Kimura, "Influence of heat flow control on dynamical spin injection in CoFeB/Pt/CoFeB trilayer," Sci. Rep. **12**, 3467 (2022).

[42] S. Martin-Rio, C. Frontera, A. Pomar, L. Balcells, and B. Martinez, "Suppression of spin rectification effects in spin pumping experiments," Sci. Rep. **12**, 224 (2022).

[43] M.A.W. Schoen, D. Thonig, M.L. Schneider, T.J. Silva, H.T. Nembach, O. Eriksson, O. Karis, and J.M. Shaw, "Ultra-low magnetic damping of a metallic ferromagnet," Nat. Phys. **12**, 839 (2016).

[44] E. Shikoh, K. Ando, K. Kubo, E. Saitoh, T. Shinjo, and M. Shiraishi, "Spin-Pump-Induced Spin Transport in p-Type Si at Room Temperature," Phys. Rev. Lett. **110**, 127201 (2013).

[45] S. Wang, T. Gao, C. Wang, and J. He, "Studies of anisotropic magnetoresistance and magnetic property of $Ni_{81}Fe_{19}$ ultra-thin films with the lower base vacuum," J. Alloys Compd. **554**, 405 (2013).

[46] H. Wang, C. Du, P.C. Hammel, and F. Yang, "Spin transport in antiferromagnetic insulators mediated by magnetic correlations," Phys. Rev. B **91**, 220410 (2015).

[47] H. Qiu, T.S. Seifert, L. Huang, Y. Zhou, Z. Kašpar, C. Zhang, J. Wu, K. Fan, Q. Zhang, D. Wu, T. Kampfrath, C. Song, B. Jin, J. Chen, and P. Wu, "Terahertz Spin Current Dynamics in Antiferromagnetic Hematite," Advanced Science **10**, 2300512 (2023).